\documentclass[mathleft
]{an}
\usepackage{graphicx}
\usepackage{times}
\usepackage{natbib}
\bibpunct{(}{)}{;}{a}{}{,}
\usepackage{aas_macros}

\begin{document}

% The following seven commands are intended for editorial usage and should be ignored by
% the author(s).
%\Pagespan{789}{}% Document's page range. 
% If second parameter is left empty, the last page is computed automatically.
%\Yearpublication{2006}%
%\Yearsubmission{2005}%
%\Month{11}%   
%\Volume{999}%  
%\Issue{88}% 
% \DOI{This.is/not.aDOI}% 

\title{A wider audience: turning VLBI into a survey instrument}

\author{E. Middelberg\inst{1}\fnmsep\thanks{Corresponding author:
  \email{middelberg@astro.rub.de}\newline}
\and  A. T. Deller\inst{2}
\and  W. F. Brisken\inst{3}
\and  J. S. Morgan\inst{4}
\and  R. P. Norris\inst{5}
}

\titlerunning{Turning VLBI into a survey instrument}
\authorrunning{Middelberg et al.}
\institute{
Astronomisches Institut der Ruhr-Universit\"at Bochum, Universit\"atsstra\ss e 150, 44801 Bochum, Germany
\and 
ASTRON, Oude Hoogeveensedijk 4, 7991 PD Dwingeloo, The Netherlands
\and 
National Radio Astronomy Observatory, PO Box O, Socorro, NM 87801
\and
International Centre for Radio Astronomy Research, Curtin University, GPO Box U1987, Perth, WA, Australia
\and
CSIRO Astronomy and Space Science, Australia Telescope National Facility, PO Box 76, Epping NSW 1710, Australia
}

\received{...}
\accepted{...}
\publonline{later}


\abstract{Radio observations using the Very Long Baseline
  Interferometry (VLBI) technique typically have fields of view of
  only a few arcseconds, due to the computational problems inherent in
  imaging larger fields. Furthermore, sensitivity limitations restrict
  observations to very compact and bright objects, which are few and
  far between on the sky. Thus, while most branches of observational
  astronomy can carry out sensitive, wide-field surveys, VLBI
  observations are limited to targeted observations of carefully
  selected objects. However, recent advances in technology have made
  it possible to carry out the computations required to target
  hundreds of sources simultaneously. Furthermore, sensitivity
  upgrades have dramatically increased the number of objects
  accessible to VLBI observations. The combination of these two
  developments have enhanced the survey capabilities of VLBI
  observations such that it is now possible to observe (almost) any
  point in the sky with milli-arcsecond resolution. In this talk I
  review the development of wide-field VLBI, which has made
  significant progress over the last three years.}

\maketitle

\section{Introduction}

Since the invention of the VLBI technique in the 1960s, observations
using it have almost exclusively been limited to the study of
carefully selected, small samples of objects. There are two
fundamental reasons which have led to this situation.  First, the
observing bandwidth is restricted by the need to record the raw data
on tape or disk and then replay and process the data afterwards on a
reasonable timescale. Since the sensitivity of an interferometer is
linked to the bandwidth of the observations (i.e., the number of
measurements being made), this limitation directly affected the
sensitivity. Connected-element interferometers such as the
VLA\footnote{Very Large Array} and ATCA\footnote{Australia Telescope
  Compact Array} are not affected by this, since there is no need to
store the raw antenna signals before the correlation is carried
out. Second, the field of view of a radio interferometer is limited by
the spectral and temporal resolution with which the data are
correlated (which can also be expressed as the channel width,
$\Delta\nu$, and integration time, $\Delta\tau$, over which the data
are averaged). In an often-used analogy the limitation arising from
bandwidth averaging can be compared to chromatic aberration in a lens,
and the limitation from time averaging to motion blur in a photograph
when the exposure time has been too long. Whilst both effects also
affect connected-element interferometers, the long baselines and
consequently high spatial resolution of VLBI observations limit the
field of view to around one arcsecond. Therefore traditional VLBI
observations are helplessly unsuitable for surveys of large fractions
of the sky.

\section{Bandwidth and time averaging smearing}

\begin{figure*}[htpb!]
\center
\includegraphics[width=\linewidth]{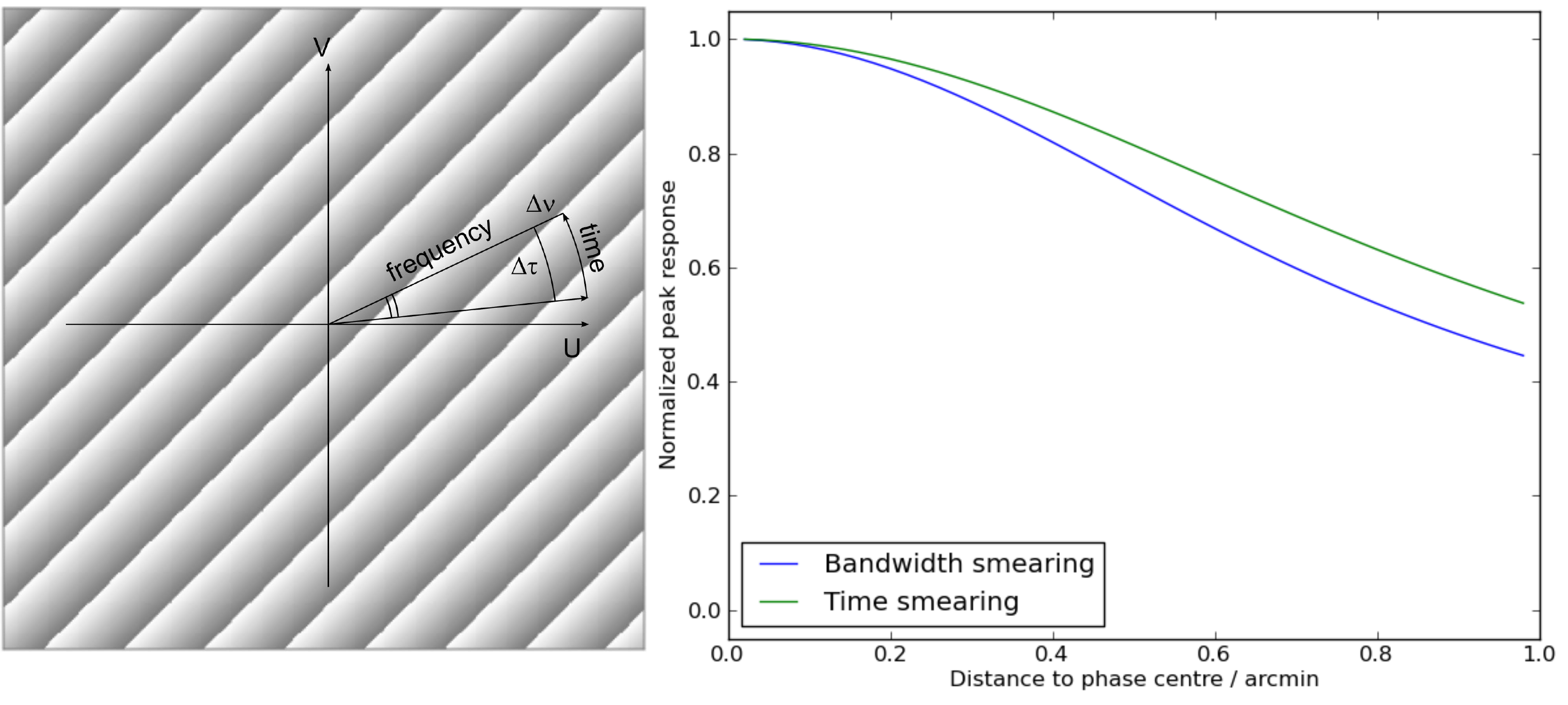}
\caption[The effects of bandwidth and time averaging]{Illustration of
  bandwidth and time smearing. {\it Left panel:} Shown is the $(u,v)$
  plane in which an interferometer baseline makes a measurement. The
  sinusoidal phase variations of a point source offset from the field
  centre are indicated by the background greyscale image. The
  frequency of the phase variations is proportional to the distance of
  a source to the field centre, and its orientation by its angle in
  the $(u,v)$ plane. The ratio of the observing frequency, $\nu$, and
  the bandwidth, $\Delta\nu$, is independent of baseline length, and
  therefore the fraction covered by $\Delta\nu$ along the baseline
  vector increases with the baseline length. The area covered by the
  baseline vector within the integration time, $\Delta\tau$, is caused
  by the angular velocity of the earth's rotation and therefore also
  is a function of distance to the centre of the $(u,v)$ plane. The
  Fourier transform of the sky brightness does not change much on
  short baselines, close to the field centre, and so the measurement
  is coherent in frequency and time. On longer baselines, however, the
  measurement is integrated over an area in which the phase of the
  Fourier transform change by more than one radian, and therefore
  coherence is lost. {\it Right panel:} the decrease of sensitivity
  caused by bandwidth and time smearing, estimated using the
  approximations given by \cite{Thompson2001}. The calculation has
  been carried out for conditions typical of traditional VLBI
  observations: a frequency of 1.4\,GHz, a channel width of 500\,kHz,
  a baseline length of 5000\,km and an integration time of 4\,s. At a
  distance of 0.4\,arcmin (24\,arcsec) both bandwidth and time
  smearing alone cause the sensitivity to decrease to about 0.84,
  causing a combined loss of 0.84$^2=0.7$. To compensate for this
  sensitivity loss the total observing time during the observations
  would need to be approximately doubled.}
\label{fig:wide-field}
\end{figure*}

In a radio interferometer image suffering from too low spectral
resolution and/or too long integration times, radio sources away from
the field centre get blurred and extended. These two effects are
called bandwidth and time smearing respectively. The magnitude of the
effect can be estimated following \cite{Thompson2001},
chap. 6.4. Here, I provide a qualitative explanation for the effects.

A radio interferometer is sensitive to the spatial coherence of the
electromagnetic radiation received at two antennas. One can show that
this spatial coherence function is, within certain limits, equivalent
to the Fourier transform of the sky brightness distribution (see
\citealt{Clark1999} for an introduction). At any one time, a single
baseline makes a measurement of the complex Fourier plane, at a
location given by the length and relative orientation of the baseline
with respect to the direction to the source. Since the sky plane has
axes with units of angles, the Fourier plane, called $(u,v)$ plane by
radio astronomers, also must have dimensionless units, which in this
case correspond to wavelengths. The axes of the Fourier plane are
chosen such that $v$ points towards the coordinate north pole.

Since a measurement requires to collect energy, it must be carried out
using a finite bandwidth and a finite integration time. Thus a
measurement carried out by a baseline does not correspond to a point
in the Fourier plane, but to a small area along an arc traced out by
the baseline in the $(u,v)$ plane (caused by the rotation of the earth
during the integration). In the case of a point source offset from the
field centre the phase in the Fourier plane changes as illustrated in
Fig.~\ref{fig:wide-field} (the amplitude is constant across the
$(u,v)$ plane for point sources). Consequently, longer baselines
measure Fourier components farther away from the field centre, and
potentially integrate over an area so large that the phase of the
Fourier transform changes, and so coherence is lost. The frequency of
the sinusoidal phase variations indicated in Fig.~\ref{fig:wide-field}
is proportional to the distance of the point source to the field
centre. Hence shorter baselines can image larger fields, and this is
the fundamental reason for the tiny fields of view of traditional VLBI
observations. An example for the magnitude of the effect is given in
the right panel of Fig.~\ref{fig:wide-field}.

\section{Enabling wide-field VLBI with software correlators}

From the illustration in the left panel of Fig.~\ref{fig:wide-field}
one can (correctly) conclude that to preserve coherence on long
baselines for wide fields of view it is sufficient to decrease
$\Delta\nu$ and $\Delta\tau$. However, increasing the field of view of
a VLBI array to a dimension which is comparable to the field of view
of the interferometer elements (of order $\lambda/D$, or 30\,arcmin)
requires to produce data with kHz resolution, and a temporal
resolution of order tens of ms. This requirement is far in excess of
what had been available at correlators until very recently.

Fortunately, the power of commodity hardware computers had increased
by 2005 to such a degree that the correlation of interferometer data
using software on computer clusters became feasible. Correlation is
what is known as an ``embarrassingly parallel'' problem, since it can
be divided up in frequency, time, and baseline, and no communication
between the correlator processes is required. In 2007 a software
correlator was presented by \cite{Deller2007}. Its main features are
almost unlimited spectral and temporal resolution, superb scalability,
and speed, since it makes use of specific optimisations in Intel
CPUs. Meanwhile a deployment of DiFX, called VLBA-DiFX, has replaced
the hardware correlator of the VLBA\footnote{Very Long Baseline
  Array}.

It was therefore feasible by 2006 to start a project to image all
known radio sources in the CDFS\footnote{Chandra Deep Field South}
with the VLBA. The observations were carried out in 2007, and the
telescope data were shipped to the MPIfR\footnote{Max-Planck Institut
  f\"ur Radioastronomie} in Bonn, where a computer cluster was
available for correlation.

\section{Wide-field VLBI observations of the CDFS}

When this project was initiated it was not clear in what way the DiFX
software correlator would be used to produce the desired data. It was
nevertheless attempted to select a field for the observations with
good complementary data to enhance the interpretative value of the
results. The CDFS is arguably the best-studied area in the sky and was
therefore chosen, despite its declination of $-27^\circ$ which makes
it a challenging target for the VLBA. The project is described in full
detail in \cite{Middelberg2011a}, and we give here a condensed summary.

\subsection{Observations, correlation and calibration}

\begin{figure*}[htpb!]
\center
\includegraphics[width=0.45\linewidth]{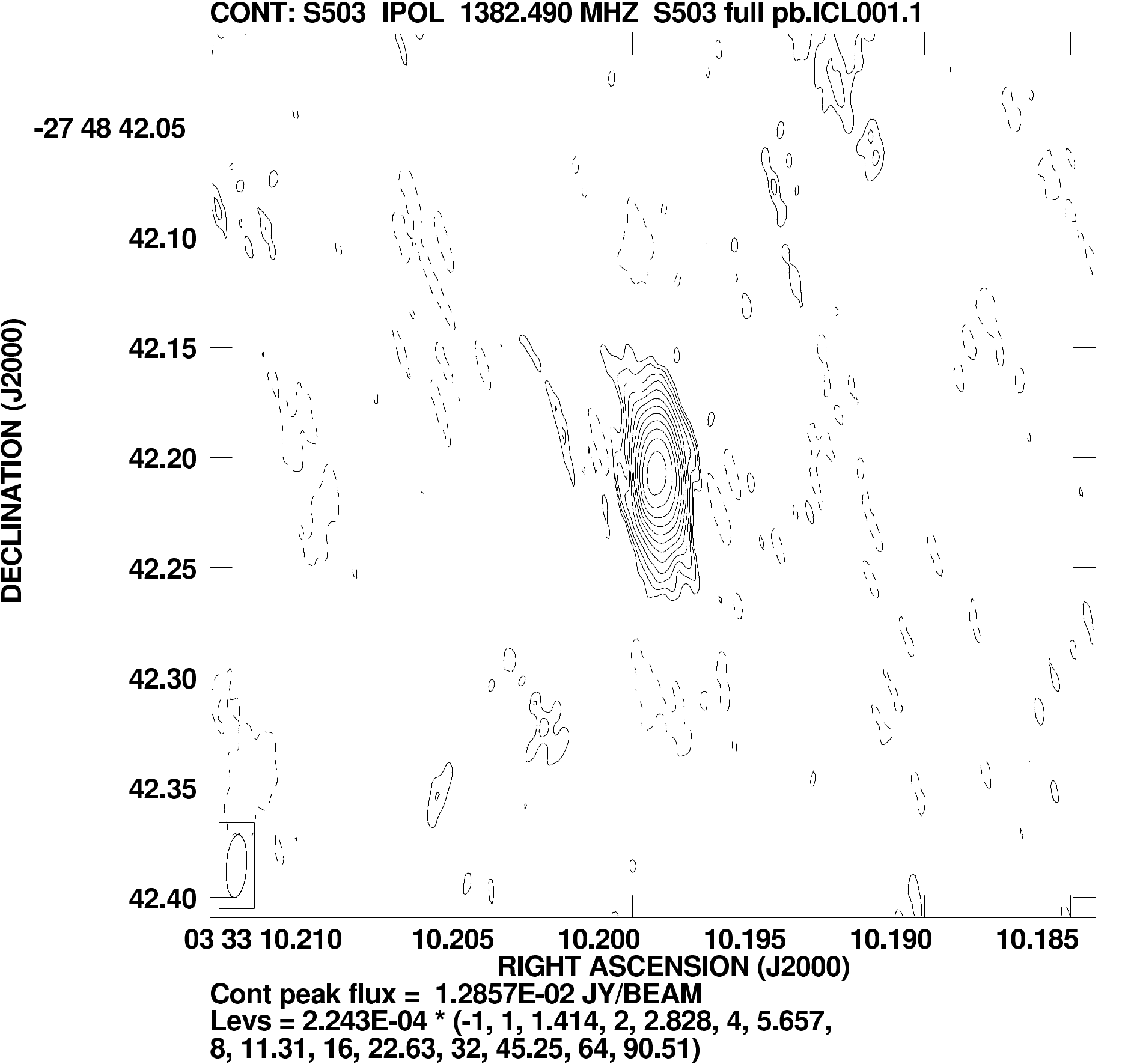}
\includegraphics[width=0.45\linewidth]{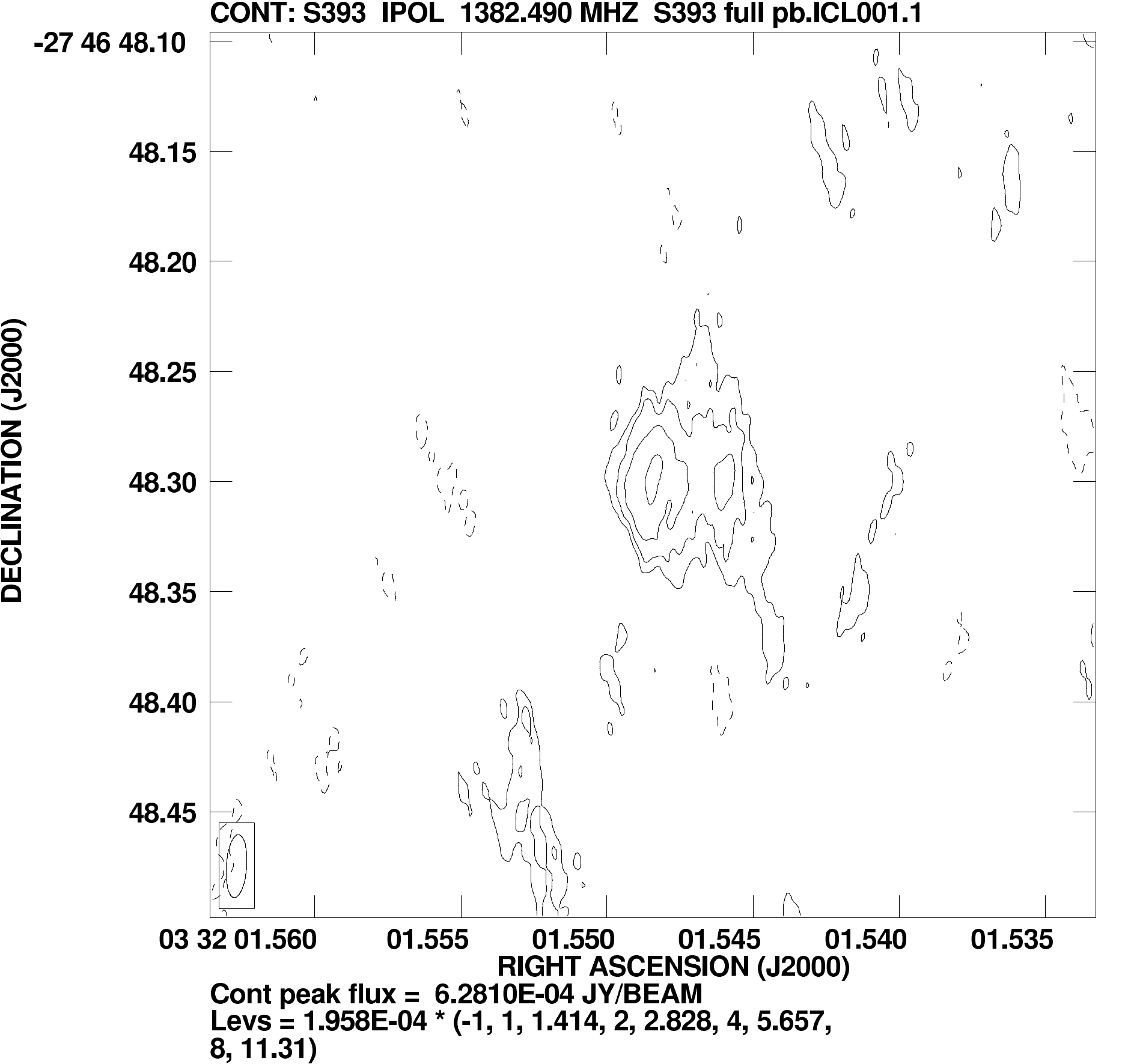}
\caption{Contour plots of two sources from the VLBI survey of the
  CDFS. Contours start at three times the image rms level and increase
  by factors of $\sqrt{2}$. Both images suffer from calibration
  artifacts, arising from the low declination of the field. {\it Left
    panel:} Contour plot of S503, the strongest detected target, which
  served as an in-beam calibrator to improve the coherence of the data
  after initial phase calibration. The image noise is
  0.75\,mJy\,beam$^{-1}$ and the integrated flux density is
  21.2\,mJy. {\it Right panel:} Contour plot of S393, one of the few
  detected targets with sub-structure on mas scales. The image noise
  is 0.65\,mJy\,beam$^{-1}$ and the integrated flux density is
  24.7\,mJy.}
\label{fig:examples}
\end{figure*}

\begin{figure}[htpb!]
\center
\includegraphics[width=0.85\linewidth,angle=270]{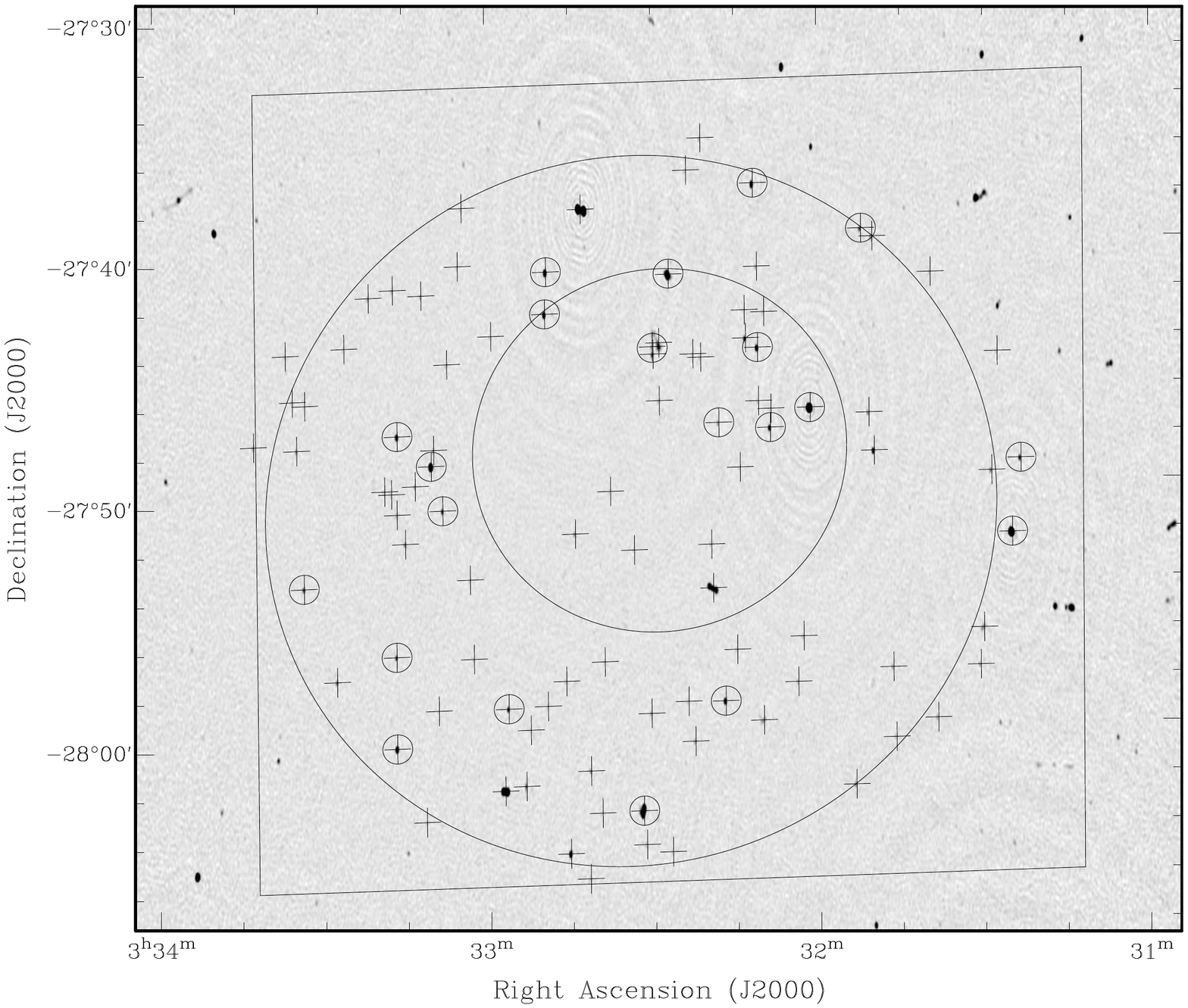}
\caption{An overview of the observed area. The background image is a
  radio image of the CDFS made with the Australia Telescope Compact
  Array (\citealt{Norris2006a}). The rms of the image is around
  20\,$\mu$Jy\,beam$^{-1}$, and the faintest sources have flux
  densities of 100\,$\mu$Jy. The square indicates the eCDFS area
  observed with the Chandra satellite by \cite{Lehmer2005} with an
  integration time of 240\,ks; the large circle indicates a typical
  VLBA antenna's primary beam size at the half-power level at
  1.4\,GHz; and the medium circle is the region covered uniformly with
  a 2\,Ms exposure with Chandra.  Crosses are drawn at the locations
  of the 96 targets taken from \cite{Norris2006a} and small circles
  around those targets which were detected with the VLBA.}
\label{fig:overview}
\end{figure}

The CDFS was observed on 3 July 2007 with the VLBA at 1383\,MHz. An
overview of the field is given in Fig.~\ref{fig:overview}. The low
declination of the field resulted in an elapsed time of the
observations of only 9\,h, and the expected sensitivity was
32\,$\mu$Jy\,beam$^{-1}$. Two effects increase this estimate: the need
to self-calibrate on a suitably bright target in the field resulted in
a loss of about 40\,\% of the data, and the low declination caused the
antenna system temperatures to increase of the order of 30\,\% to
40\,\%. The final sensitivity in the centre of the field was found to
be 55\,$\mu$Jy\,beam$^{-1}$.

To keep the amplitudes within 5\,\% of their true values, one would
have to use a channel width of 4\,kHz and 50\,ms integrations. Such a
correlation would result in around 3\,TB of visibility data, which
would be very difficult to manage even on large general-purpose
computers. Fortunately, the DiFX software correlator has recently been
extended to support the simultaneous correlation of multiple field
centres. This mode is analogous to what has previously been known as
``multi-pass correlations'' on hardware correlators, but in DiFX it is
implemented efficiently and supports several hundred phase centres at
a comparatively small computational cost. The details of this mode are
described in \cite{Deller2011}. The idea is to place one phase centre
on each known radio source. The efficiency of the DiFX implementation
arises from the fact that the computationally expensive FFT of the
incoming antenna data needs to be computed only once, and the
subsequent computations for each phase centre only require complex
multiplications at a rate of around 100 per second. This procedure
results in one normal-sized VLBI data set per phase centre. The field
of view of each data set is small (of order 10\,arcsec), but the
source coordinates are sufficiently well known that bandwidth and time
averaging effects are not an issue.

Amplitude calibration was carried out using $T_{\rm sys}$ measurements
and known gain curves. Initial phase calibration was done using
regular interleaved scans on a nearby calibrator source. However, the
low declination of the field caused significant incoherence of the
data and required self-calibration. Fortunately one target was
detected with sufficient SNR for self-calibration, although
visibilities on long baselines had poor SNR and caused a loss of
around 40\,\% of the data which could not be calibrated. Since the
phase response of a VLBA antenna is constant within the primary beam,
and since the geometric delays had been compensated at the correlation
stage, the amplitude, phase, and delay correction could simply be
copied from one target source to another, and images could be
formed. However, another correction is needed to compensate for the
amplitude response arising from primary beam attenuation, which
reduces the apparent flux density of a source by up to 50\,\%.

Unlike in compact-array interferometry, where the primary beam
correction is carried out in the image plane, we corrected for primary
beam attenuation by calculating visibility gains. This is possible
only because the fields of view are very small in our observations and
so the attenuation due to the primary beam does not vary significantly
across the image. We have developed a correction scheme which uses
measured beam widths and the relative orientations of the antennas and
targets in the telescopes' field of views to compensate for the
primary beam attenuation. As a result we were able to produce
calibrated 1.4\,GHz VLBI images for 96 targets from a single observing
run with the VLBA.

\subsection{Results}

Out of the 96 targets, 20 were confidently detected (peak flux density
exceeding 6 times the local noise), and one more target was detected
with lower significance. Two examples are shown in
Fig.~\ref{fig:examples}. The large amount of ancillary data in the
CDFS allowed us to supply our observations with substantial
multi-wavelength data. In particular, a redshift was available for
almost all sources, and the entire field had been covered with deep
X-ray observations as part of the CDFS and the eCDFS (``extended''
CDFS) surveys. It was found that despite the large amount of data,
seven of the detected targets had not previously been identified as
harbouring an active nucleus. Since our VLBI detections required
brightness temperatures in excess of $4.2\times10^5$\,K to make a
detection, each of these sources could readily be identified as
containing an AGN (using the available redshifts for the additional
argument that the luminosity must exceed that of galactic,
high-brightness temperature sources). The multi-wavelength data of one
source, S423, indicates that it is a starburst galaxy, however, the
VLBI detection clearly shows that it also contains an AGN. Combined
sources such as this are rare, but they are thought to be more
frequent at high redshifts, and to identify them in large numbers
requires VLBI survey techniques. A marginal detection was made in
S443, the location of which at the edge of a spiral galaxy (visible in
a {\it Hubble} Space Telescope image) suggests that this could be a
radio supernova.

\section{Wide-field VLBI observations of the Lockman Hole/East region}

\begin{figure}[htpb!]
\center
\includegraphics[width=0.85\linewidth]{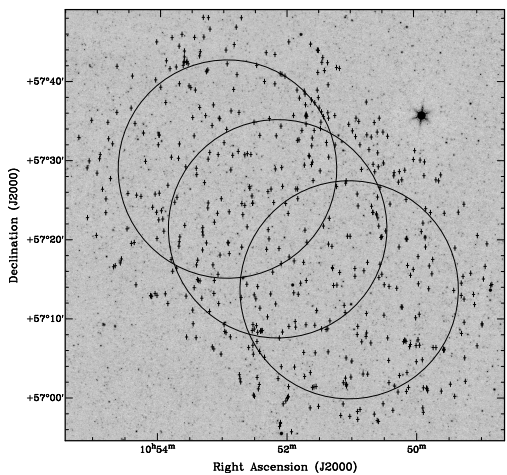}
\caption{Overview of the VLBA observations of the Lockman
  Hole/East. Shown is a {\it Spitzer} 3.6\,$\mu$m greyscale image,
  circles denote the typical FWHM circle of a VLBA antenna at
  1.4\,GHz, and crosses indicate the locations of the 532 targets.}
\label{fig:bm332-overview}
\end{figure}

The observations described in the previous section were mainly
intended to explore the capabilities of the DiFX correlator in
wide-field VLBI observations, and to develop a primary beam
calibration scheme. For larger surveys, as they are commonly carried
out in mosaic mode with compact interferometers, however, additional
techniques were needed. For example, only the fortuitous circumstance
of having a suitably bright target within the primary beam allowed us
to calibrate (most of) the data. In a survey of a larger region with
many overlapping pointings, not many such sources can be expected,
preventing calibration. A project was therefore initiated to take
wide-field VLBI one step further, and the Lockman Hole/East region was
chosen as the target. This field had been observed with the VLA, with
a sensitivity of less than 10\,$\mu$Jy\,beam$^{-1}$, resulting in the
detection of 1450 radio sources. Three full synthesis observations of
12\,h duration were needed to cover the area. Since the sensitivity of
the VLBA reaches around 24\,$\mu$Jy\,beam$^{-1}$ in this time, the
overlap of the three pointings resulted in a sensitivity of around
18\,$\mu$Jy\,beam$^{-1}$, and all 532 sources with integrated flux
densities of more than 100\,$\mu$Jy were targeted. An overview of this
field is shown in Fig.~\ref{fig:bm332-overview}.

\subsection{Observations and calibration}

In 2010, three overlapping pointings were observed. Each pointing
contained around 320 suitable targets, and each of these targets were
used for correlation using the new DiFX multi-phase centre mode. Many
targets were present in more than one pointing, and only the
combination of these pointings would result in the full available
sensitivity.

\begin{figure*}
\center
\includegraphics[width=0.48\linewidth]{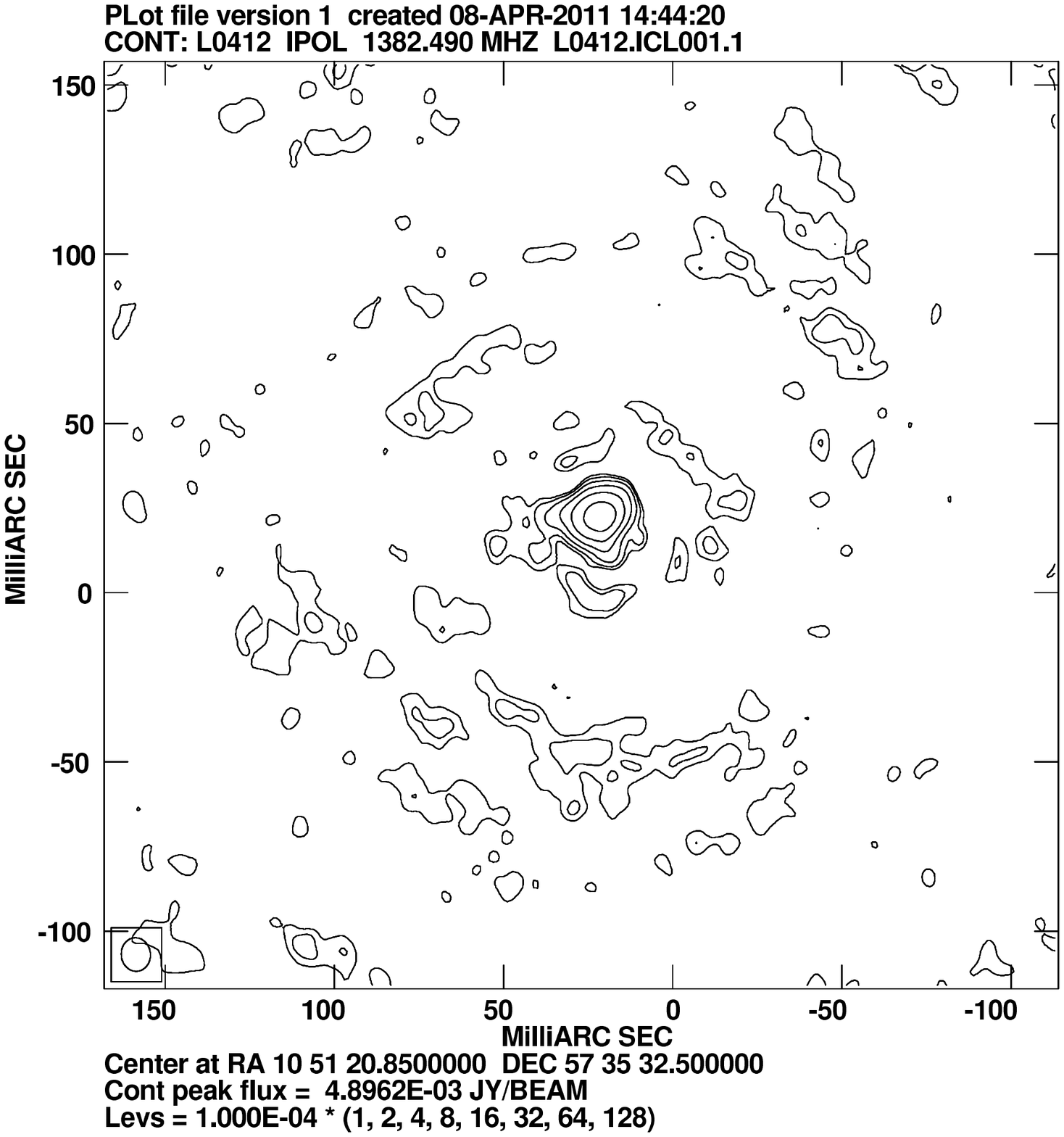}
\includegraphics[width=0.48\linewidth]{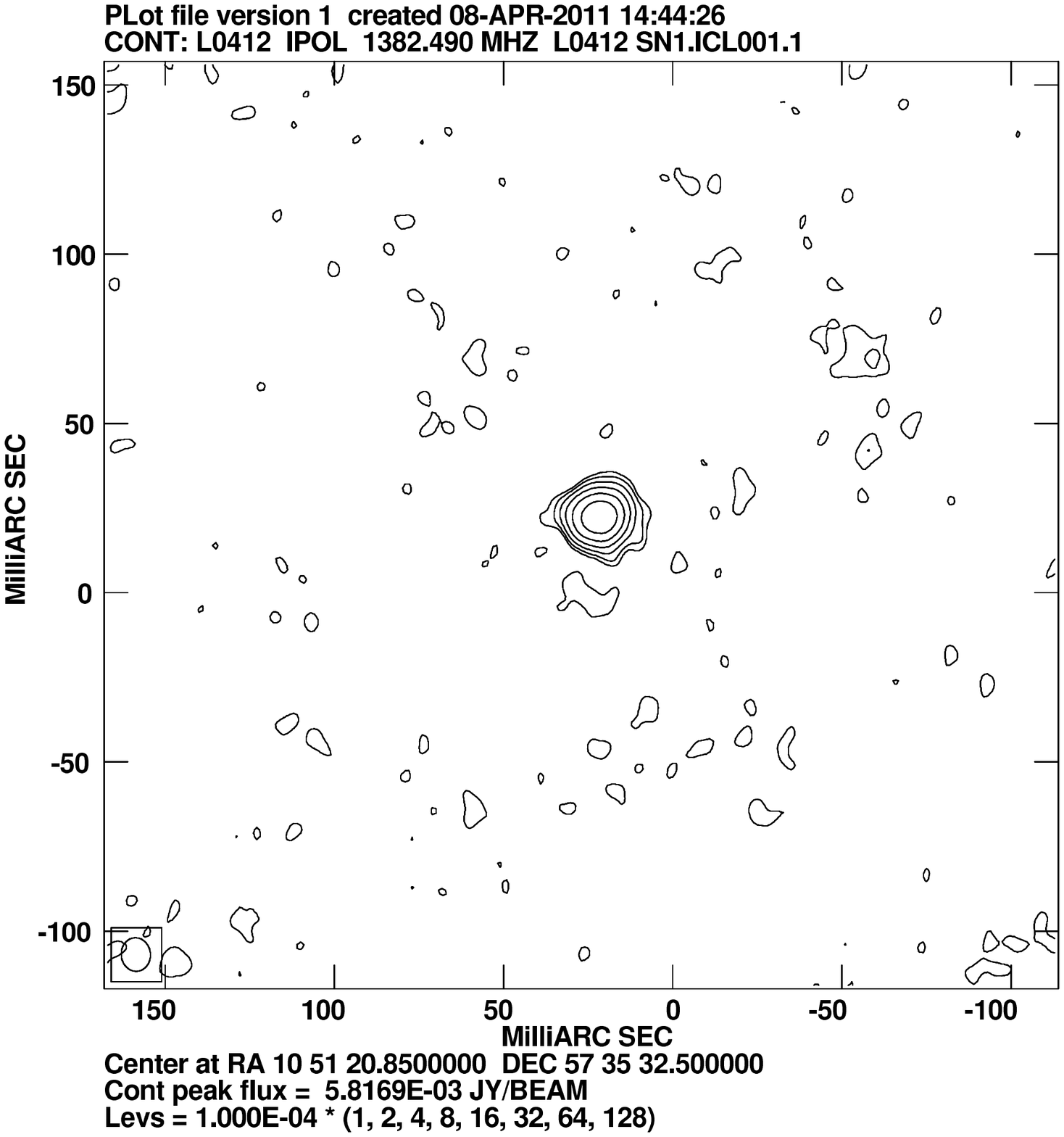}
\caption{Contour plot of the radio source L0412 before (left) and
  after (right) multi-source self-calibration has been
  applied. Contours are drawn at 0.1\,mJy$\times$(1, 2, 4, ...). The
  signal-to-noise ratio was improved from from 72 to 115. The peak
  flux density has increased significantly, the noise has dropped, and
  image artifacts are much reduced. }
\label{fig:multi-source-selfcal}
\end{figure*}

After initial calibration no targets were found sufficiently bright
for self-calibration, and a calibration scheme had to be developed to
use the signal from many faint, separate sources for
self-calibration. This scheme works as follows:

\begin{itemize}
\item Phase-referenced images were made for all targets, and the
  images were searched for emission. Typically around 30 sources were
  found with a SNR of more than 7, with the brightest reaching an SNR
  of almost 100. Since the data are later combined using weights which
  are proportional to the square of the SNR, a source with SNR=10
  contributes only 1/100 of the combined signal compared to a source
  with SNR=100. Therefore only the brightest 10 sources or so were
  used in the following steps.

\item The individual data sets were divided by the Clean model
  obtained during imaging. This results in data sets each showing a
  1\,Jy point source in the field centre. It is worth noting that in
  this process the data weights are modified according to the
  amplitude of the input model. Since at this stage no primary beam
  correction had been carried out, the noise in all images is the
  same, and so the measured amplitude is directly proportional to the
  SNR in the image. Hence this process conveniently takes care of
  proper weighting when the data are combined.

\item The source coordinates in the data set headers were set to the
  same value. This results in data sets with a 1\,Jy source at the
  same coordinates, even though the signal in the data comes from
  separate sources. For each baseline, time, and frequency the
  combined data set now contains multiple measurements of a point
  source.

\item Self-calibration was used with a 1\,Jy point source model to
  improve the coherence of the combined data set. The phase
  corrections derived in this process were then copied to all original
  data sets, and improved images could be made. The improvement in
  image quality is illustrated in Fig.~\ref{fig:multi-source-selfcal}.
\end{itemize}

As a result noise-limited images of all targets from each epoch were
obtained. A fundamental consequence of this procedure, however, should
not go unnoticed: it substantially lowers the sensitivity threshold
for self-calibration. Therefore, given sufficient instantaneous
sensitivity and large primary fields of view, most of the sky is
accessible to VLBI observations at low GHz frequencies. All it takes
to calibrate the data is to add a few (tens) of phase centres towards
known sources, taken for example from the NVSS\footnote{National Radio
  Astronomy Observatory Very Large Array Sky Survey} survey. These
data can then be used in multi-source self-calibration and the phase
correction can be applied to the target source data.

Another important step was to combine the data from the same target
observed during different epochs. Since the positions of the targets
were essentially referenced to the position of the calibrator used
during the observations, it is expected that a systematic offset can
be found between epochs. This offset was determined by measuring the
peak positions of several bright, detected targets in each epoch and
calculating the average of the position differences. The systematic
offsets were found to be small, of order less than 1\,mas
(approximately 1/10 of a synthesized beam width). The data were then
phase-shifted to compensate for this offset. Subsequently, primary
beam calibration as described in the previous section was performed
and the data for each source, taken from the different epochs, were
combined and imaged.

\subsection{Results}

The calibration of the data had not been finished at the time of
writing. In particular it was noticed that the primary beam correction
scheme was unsuitable when sources are located outside the primary
beam (which is the case for a substantial number of targets, see
Fig.~\ref{fig:bm332-overview}.) More accurate beam measurements are
underway, and calibration is expected to be finished by the end of
2011. A preliminary analysis, however, indicates that around 70
targets have been detected, a fraction smaller than in the CDFS. One
effect leading to this result probably is that the sources are
generally fainter, and that therefore more starburst galaxies are
contained in the sample, which leads to fewer detections.

\section{Conclusions}

Wide-field VLBI observations are now practical and relatively easy to
carry out. The new multi-phase centre mode of the DiFX software
correlator allows one to observe a large number of sources
simultaneously in a single observation, and the multi-source
self-calibration strategy we present here enables noise-limited
calibration of these data. We point out that wide-field VLBI
observations have a number of applications outside the study of galaxy
populations in sensitive surveys: a sample of (relatively) faint
background sources can serve as a reference in astrometric
observations to measure proper motions and parallaxes; and
observations of faint targets are no longer limited by the accuracy of
phase corrections derived from interleaved calibrator
observations. The full potential of wide-field VLBI observations,
however, is expected to be reached in even more sensitive observations
of large fields such as the COSMOS survey, or in less sensitive
observations of very large areas (tens or hundreds of square
degrees). 

\acknowledgements

\newpage%%%%%%%%%%%%%%%%%%%%%%%%%%%%%%%%%%%%%%%%%%%%%%%%%%%%%%

\end{document}